\documentclass[review]{elsarticle}

\usepackage{lineno,hyperref}
\modulolinenumbers[5]

\journal{Journal of \LaTeX\ Templates}

\begin{document}

\begin{frontmatter}

\title{On a possible fractal relationship between the Hurst exponent and the nonextensive Gutenberg-Richter index}

\author{D. B. de Freitas} 
	\address{ Departamento de F\'{\i}sica, Universidade Federal do Cear\'a, Caixa Postal 6030, Campus do Pici, 60455-900 Fortaleza, Cear\'a, Brasil}
	\ead{danielbrito@fisica.ufc.br} 
\author{G. S. Fran\c{c}a}
	\address{ Observat\'orio Sismol\'ogico-IG/UnB, Campus Universit\'ario Darcy Ribeiro SG 13 Asa Norte, 70910-900 Bras\'{\i}lia, Brasil}
\author{T. M. Scheerer}
	\address{ Conselho Nacional de Desenvolvimento Cient\'ifico e Tecnol\'ogico, CNPq, Brazil}
\author{C. S. Vilar}
	\address{ Instituto de F\'{\i}sica, Universidade Federal da Bahia, Campus Universit\'ario de Ondina, 40210-340 Salvador, Brasil}
\author{R. Silva}
	\address{ Departamento de F\'{\i}sica,
	Universidade Federal do Rio
	Grande do Norte, 59072-970, Natal,  RN, Brasil} 

\begin{abstract}
In the present paper, we analyze the fractal structures in magnitude time series for a set of unprecedented sample extracted from the National Earthquake Information Center (NEIC) catalog corresponding to 12 Circum-Pacific subduction zones from Chile to Kermadec. For this end, we used the classical Rescaled Range ($R/S$) analysis for estimating the long-term persistence signature derived from scaling parameter so-called Hurst exponent, $H$. As a result, we measured the referred exponent and obtained all values of $H>0.5$, indicating that a long-term memory effect exists. The main contribution of our paper, we found a possible fractal relationship between $H$ and the $b_{s}(q)$-index which emerges from nonextensive Gutenberg-Richter law as a function of the asperity, i.e., we show that the values of $H$ can be associated with the mechanism which controls the abundance of magnitude and, therefore, the level of activity of earthquakes. Finally, we concluded that dynamics associated with fragment-asperity interactions can be emphasized as a self-affine fractal phenomenon.
\end{abstract}


\end{frontmatter}


\section{Introduction}
In the vast majority, geophysical signals fluctuate in an irregular and complex way along the time, presenting inhomogeneous variations and extreme events, as such irregular rupture propagation and non-uniform distributions of rupture velocity, stress drop, and co-seismic slip \cite{telesca0}. The presence of scaling properties in geophysical data is an evidence that fractal method may provide a viable way to investigate the behavior of the fluctuations of the earthquake magnitudes \cite{li}. If, on the one hand, the terrestrial tectonic activity is due to very complex mechanisms that involve many variables such as deformation, rupture, released energy, land features, heterogeneity in seismogenic plate interface \cite{kawa,scherrer}. On the other hand, there is a range of different methods and tools that can be used for a better description of dynamical properties of earthquakes \cite{omori,gr}.

In particular, several statistical methods are available in the scientific literature, where use the concept of fractality. Among them, we can find methods based on self-similar and self-affine fractals such as the box dimension \cite{peitgen2004chaos}, the detrended fluctuation analysis (DFA) \cite{1992Natur.356..168P}, the detrending moving average analysis (DMA) \cite{ale}, the scaled windowed variance analysis (SWVA) \cite{1985PhyS...32..257M}, and so on. In this context, we decided to focus our attention on the seminal parameter proposed by Hurst \cite{hurst1951} to describe the long-term dependence of water levels in river and reservoirs \cite{seuront}. Unlike the current trend that directly applies the multifractal methods \cite{telesca1}, \cite{telesca2} and \cite{telesca3}, we decided to investigate the dynamics of earthquakes characterized by the Hurst exponent, where features as memory and long-term correlations are investigated. As quoted by Telesca \cite{telesca2}, a statistical analysis based on fractals are featured by power-laws and can be a powerful tool to examine the temporal fluctuations at different scales when applied to earthquake magnitude time series.

In the present paper, we examine the scaling properties of the geophysical time series obtained from Circum-Pacific subduction zones initially treated by Scherrer \textit{et al}. \cite{scherrer}. In this sense, our study applied the rescaled range ($R/S$) analysis as a self-affine fractal method to seismic data. de Freitas \textit{et al}. It is worth noting the universal character of the $R/S$ method in the analysis of the behavior of fluctuations. A large number of studies at different areas of knowledge has shown that the so-called Hurst exponent extracted from within the $R/S$ analysis provides a robust and powerful statistical method to characterize nonstationary fluctuations at different timescales \cite{defreitas2013}, \cite{suyal} and \cite{li}. More recently, \cite{defreitas2013a} found that for San Andreas fault the Hurst exponent of 0.87, indicated a strong long-term persistence. Other studies (e.g., \cite{li}) also indicate that the Hurst exponent of seismic data calculated by R/S method is greater than 0.6.

Our main interest is to examine a possible correlation between the scaling properties of the subduction-zone earthquakes on the Circum-Pacific controlled by the interaction of asperities \cite{lay} and the Hurst exponent estimated from the ($R/S$) analysis \cite{hurst1951,mw1969a}. In general, we believe that different subduction zones distributed in major groups (for more details, see Section 3) can be associated with distinct scaling laws which relate the dynamical state of the earthquakes and the long-term persistence. Moreover, this procedure could be used to distinguish the zones with the distribution of stronger stress from the weaker ones. As mentioned by Lay and Kanamori \cite{lay}, the interaction and failure of an asperity can cause an increase in stress on the adjacent asperities. In this context, the authors elaborate a general structure of categories based on the extreme behavior of dynamics and strength of the earthquakes as cited in Scherrer \textit{et al}. \cite{scherrer}.

Our paper is organized as follows: in next section, we describe the Hurst method used in our study and a brief discussion about the nonextensive formalism. In Section 3, we present our seismic sample. The main results and their physical implications are presented in section 4, and conclusions are summarized in the last section.

\begin{figure*}
	\includegraphics[width=0.95\textwidth]{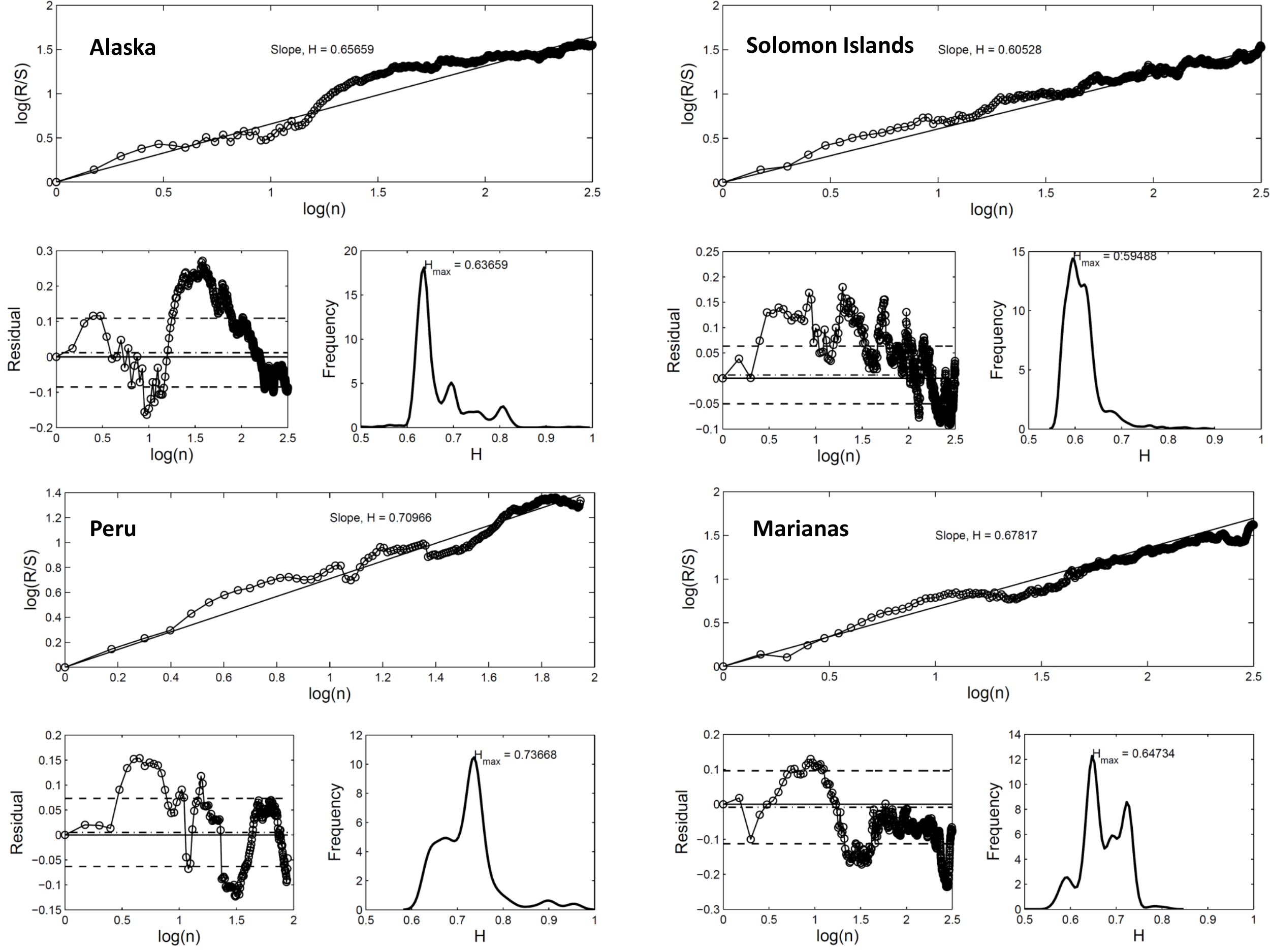}
	\caption{\textit{Top panel}: Individual $\log(R/S)$ points as a function of the logarithm of box-size $n$ for 4 subduction zones from our sample. \textit{Left bottom panel}: Residual extracted from the difference between the $\log(R/S)$ points and the best linear adjustment. The solid line represents the perfect agreement, the dashed line denotes the mean value, whereas the dash-dotted lines indicate 1$\sigma$. \textit{Right bottom panel}: This plot is the Kernel adjustment of $H$ calculated as the derivative of the $R/S$ curve after each iteration $n$. $H_{max}$ is the maximum value of the distribution of $H$ which in all cases differs slightly from the value of $H$ identified by a straight line in the top panels.}
	\label{fig1}
\end{figure*}

\begin{figure*}
	\includegraphics[width=0.95\textwidth]{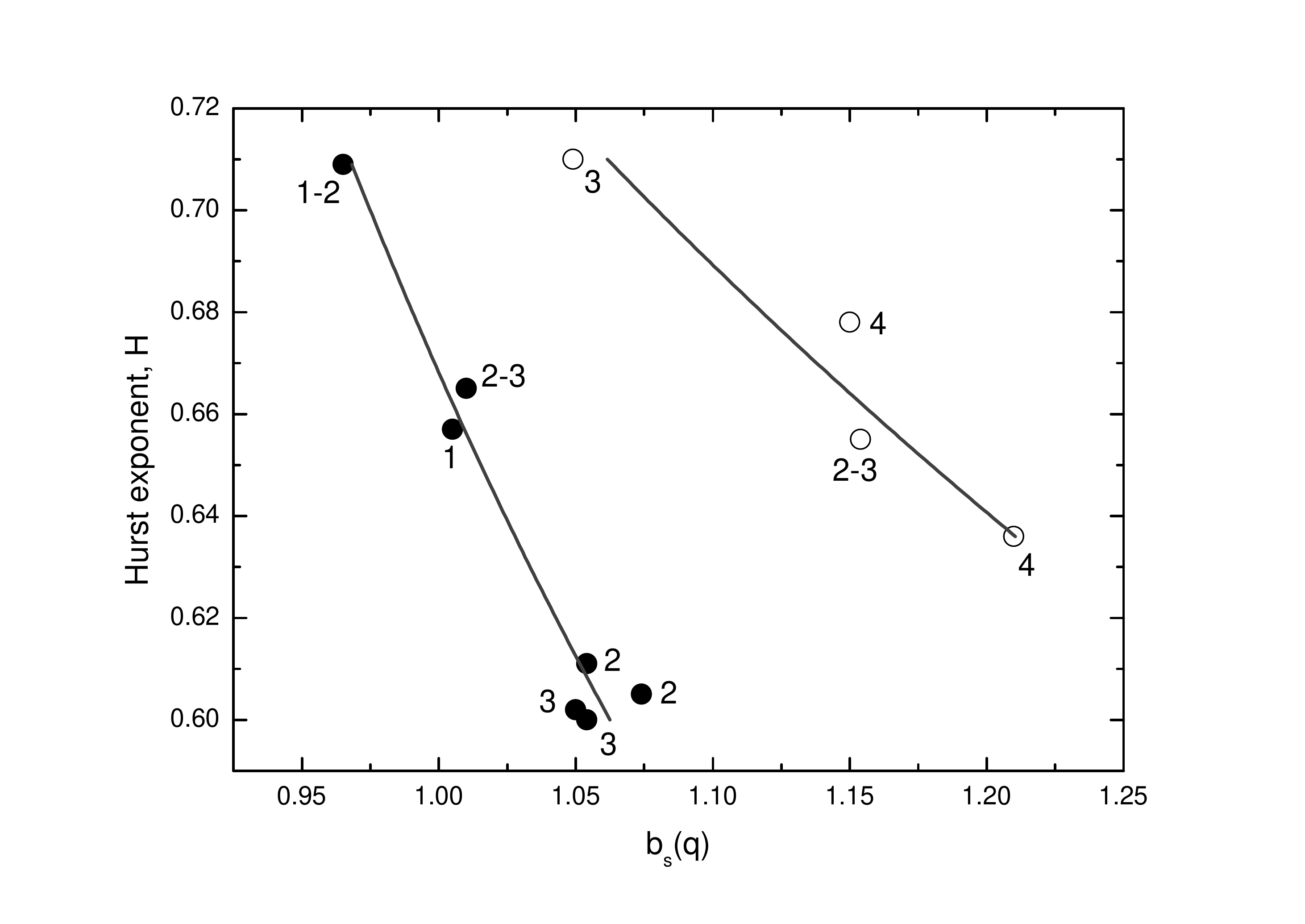}
	\caption{Values of the $b_{s}$-index extracted from the slope of the modified Gutenberg-Richter law as a function of $H$-index. Numbers indicate the category of each subduction zone (see Table 1, 2nd column and Table 2 from \cite{scherrer}). Solid and open circles are used to distinguish the two different domains.}
	\label{fig2}
\end{figure*}

\section{Statistical background}
\subsection{Hurst effect}
As pointed out by Seuront \cite{seuront}, there are various methods to describe the behavior the time series using self-affine fractals. In general, fractality traces can be estimated considering the type of signal. That signal type is limited by either noisy time series or a regular signal like a sine wave. Mandelbrot and Wallis \cite{mw1969a} and \cite{mw1969b} introduced the concept of fractional Brownian motion (\textbf{fBm}) as a generalization of Brownian motion which assume the motion of an object as a union of rescaled copies of itself uniformly distributed in all directions, i.e., a self-similar fractal. In contrast, fBm considers the rescaling of the copies of itself dependent on the direction denoted as a self-affine \cite{seuront}. As mentioned by \cite{seuront}, a series of successive increments in a fBm defines a fractional Gaussian noise (\textbf{fGn}). This other type of time series based on the increment of a fBm yields a stationary signal with mean zero. Strictly speaking, a geophysical signal can be understood as either fGn-like one or a noise-like time series \cite{defreitas2013a}.

The scientific literature points out different techniques for exploring time series of a fractal point-of-view. In general, characteristics of time series include a wider spectrum of complexity measurements due to nonstationarity, nonlinearity, fractality, stochasticity, periodicity, chaos and so on \cite{tang}. A powerful fractal technique for dealing with these assumptions is the Hurst analysis. As proposed by Hurst \cite{hurst1951}, we will focus on Rescaled Range Analysis also known as $R/S$ analysis.

\subsubsection{Fractal index measured by $R/S$ method}

As described by Hurst \cite{hurst1951}, a fractal analysis is used as a procedure to measure the long-term memory or correlation of a time series. The method developed by Hurst \cite{hurst1951}, known as the rescaled range analysis (hereafter $R/S$ analysis), is the method adopted here to estimate the Hurst exponent $H$. Our aim is to verify the capability of the cited method to distinguish the different properties in the subduction zones through the behavior of $H$ and its possible correlation with the Gutenberg-Richter indexes measured by Scherrer \textit{et al}. \cite{scherrer}.

Firstly, we consider a time series given by $x(t):=x(1),x(2),...,x(N)$ with a time-window of length $N$. Following the same procedure mentioned by de Freitas \textit{et al}. \cite{defreitas2013}, the analysis starts with two elements, $n = 2$ and for each iteration one more element is added up to $n=N$ (i.e., the whole series). In each cumulative window, we can measure two quantities denoted by $R(n)$ (the distance between the minimum and the maximum value of the accumulated deviations from mean of $x(t)$ within the window $n$) and $S(n)$ (the standard deviation of the values of $x(t)$ in this same time window). In particular, the parameter $R(n)$ is measured over a trend in the window, where it is calculated as the straight line between the first and the last points. From a mathematical point-view, $R(n)$ can be written as
\begin{equation}
\label{eq1}
R(n)=max_{1\leq t\leq n}X(t,n)-min_{1\leq t\leq n}X(t,n),
\end{equation}
where the variable $X(t,n)$ is defined as $X(t,n)=\sum^{n}_{t=1}\left(x(t)-\left\langle x(t)\right\rangle_{n}\right)$. Therefore, the long-trend is removed by the mean over the values of $x(t)$ within the window. On the other hand, the variable $S(n)$ is defined as
\begin{equation}
\label{eq2}
S(n)=\left[\frac{1}{n}\sum^{n}_{t=1}\left(x(t)-\left\langle x(t)\right\rangle_{n}\right)^{2}\right]^{1/2}.
\end{equation}

Based on these two parameters, we define the $R/S$ statistics of the fluctuations in time series as a power-law dependence over a box of $n$ elements given by:
\begin{equation}
\label{eq3}
\frac{R(n)}{S(n)}=kn^{H},
\end{equation}
where $k$ is a constant and the Hurst index ($H$) can be measured by fitting the slope of the log-log plot of $R(n)/S(n)$ versus $n$ based on least squares estimation given by
\begin{equation}
\label{eq3a}
\log\left[\frac{R(n)}{S(n)}\right]\sim H\log(n).
\end{equation}

A time series described by the $R/S$ statistic is said to be fractal if the $R/S$ curve is a perfectly straight line; this means that the residual between this straight line and the $R/S$ curve must be null. If the $R/S$ curve presents some fluctuations (however not large) around the straight line that estimate $H$, this could be due to the fact that the data are nor ideal but observational and so affected by measurement errors that make the curve $R/S$ to depart (but not with large amplitude) from the straight line. If the $R/S$ curve does not follow a straight line or if the departures from the straight line are very large the object is not fractal.

As reported by de Freitas \textit{et al}. \cite{defreitas2013}, if $H>0.5$ indicates the presence of long memory in the time series, i.e. a persistence signal. For $H=0.5$ the time series is a ``Brownian'' process, while for $H<0.5$ the series is anti-persistence (short memory). In the latter case, the signal tends not to continue in the same direction, but to turn back on itself giving a less smooth time series \cite{hm}. Besides, the values of $H$ close to zero implies strong anti-correlations. In the geophysical scenario, earthquakes bear dual features of randomicity and regularity and, therefore, it is expected a value of $H$ between 0.5 and 1.

\subsection{Nonextensive framework}
Inspired by multifractals, Tsallis \cite{tsallis1988} proposed a generalization od=f the Boltzmann-Gibbs (BG) entropy $S_{q}$
\begin{equation}
\label{eq8}
S_{q}=-k_{B}\frac{1-\Sigma_{i=1}^{W}p^{q}_{i}}{q-1}
\end{equation}
based on the entropic index $q$ which measures the degree of nonextensivity of the system. In eq. (\ref{eq8}), $k_{B}$ denotes Boltzmann's constant, $W$ is the total nember of microscopic states and $p_{i}$ represents a set of probabilities. In the limit $q=1$, we recover the classical BG entropy. 

By using that formalism, Sotolongo-Costa and Posadas \cite{costa04} and Silva \textit{et al}.\cite{silva} developed a new approach to describe the distribution of earthquakes with magnitude larger than $m$. According to Scherrer \textit{et al}. \cite{scherrer} and firstly demonstrated in \cite{sarlis}, the $b_{s}$ is related to entropic index $q$ by expression
\begin{equation}
\label{eq5}
b_{s}=\frac{2(2-q)}{q-1}.
\end{equation}

The Gutenberg-Richter law is an asymptotic relation between the total number of earthquakes $N$ and the magnitude $m$, given by
\begin{equation}
\label{eq9}
\log(N_{m})=a+b_{GR}m.
\end{equation}

As quoted by Scherrer \textit{et al}. \cite{scherrer}, the values of $b_{GR}$ are calculted using a Software Package to Analyze Seismicity denoted by ZMAP\footnote{http://mercalli.ethz.ch/~eberhard/zmap.zip}. The authors found that the values of $b_{GR}$ differ from those estimated by a nonextensive fit. However, those indexes ($b_{s}$ and $b_{GR}$) indicate higher values of $b$ for zone 4 and lower values for zone 1(as shown in table \ref{tab1} from this paper and table 4 from \cite{scherrer}). From the geophysical point of view, the $q$-values have correlation with properties of subduction zones. In this sense, it is reasonable to think if there is a correlation between $H$s and $b_{s}$. This matter will be dealt with in section 4.

\section{The seismic data}
Scherrer \textit{et al}. \cite{scherrer} have produced a list of 14 circum-Pacific subduction zones distributed in a belt along the so-called Fire Ring (see Fig.1 from mentioned paper). These data were extracted from the National Earthquake Information Center (NEIC) catalog. From that sample, we select 12 regions for applying our analysis as shown in Table \ref{tab1}. As reported by \cite{scherrer}, the NEIC catalog offers magnitude time series in different magnitudes types ($Mw,Mb,Ms,Ml,Md$) for the same event and, therefore, we choose to follow NEIC automatic ranking. In addition, we consider that using this sequence makes no significant impact on the final result of the present paper because the differences between magnitudes types are small. 

The data sample used by Scherrer \textit{et al}. \cite{scherrer} is distributed in four different subduction zones defined by asperity and broadness of rupture front. The main structure of zones is described by these authors. In this context, the reader is referred to Scherrer \textit{et al}. \cite{scherrer} for instrumental procedure and classification.

For the present analysis, we considered a magnitude greater than 3, in this case, the only effect from macroearthquakes is analyzed. Scherrer \textit {et al}. \ \cite{scherrer} measured three important parameters: (i) the entropic index $q$ which emerges from the nonextensive statistical mechanics \cite{tsallis1988}, \cite{abe01} and \cite{gell2004}, (ii) the $b_{s}$-index extracted from the slope of the nonextensive Gutenberg-Richter law \cite{costa04} and \cite{silva}, and (iii) the classical index $b_{GR}$ from Gutenberg-Richter law \cite{gr} by ZMAP software. In the next Section, we compare the Gutenberg-Richter indexes with the Hurst exponent $H$. 

\begin{table*}
	\caption{Identifier number of $SZ$ and $b_{s}$ extracted from \cite{scherrer}, and $H$ and $\sigma^{H}_{Kernel}$ estimated by our analysis. Symbols $\circ$ and $\bullet$ are used to differentiate the subsamples shown in Fig.\ref{fig2}}
	\label{tab1}
	\begin{center}
		\begin{tabular}{lccccc}
			\hline
			Area & $SZ$ & $b_{S}$ & $H$ & $\sigma^{H}_{Kernel}$\\
			&  &  & & \\
			\hline
			Alaska $\circ$ & 1&1.005 & 0.657 & 0.16\\
			Aleutians $\circ$ & 1-2&0.965 & 0.709 & 0.15\\
			Central America $\bullet$ & 2-3&1.154 & 0.655 & 0.09\\
			Central Chile $\circ$ & 3 &1.054 & 0.600 & 0.16\\
			Colombia $\circ$ & 2 &1.054 &0.611 & 0.17\\
			Kuriles $\circ$ & 3 &1.050 & 0.602 & 0.14\\
			Marianas $\bullet$ & 4 &1.150 & 0.678 & 0.16\\
			New Hebrides $\circ$ & 2-3&1.010 & 0.665 & 0.19\\
			Peru $\bullet$ &3 &1.049 & 0.710 & 0.13\\
			Solomon Islands $\circ$ &2 &1.074 & 0.605 & 0.10\\
			Tonga \& Kermadec $\bullet$ & 4&1.210 & 0.636 & 0.13\\
			\hline
		\end{tabular}
	\end{center}
\end{table*}

\section{Results and discussions}
As shown in the top panels from Fig. \ref{fig1}, the $R/S$ method was used to estimate the values of the Hurst exponent for a data set of 12 circum-Pacific subduction zones. As a result, the values of $H$ were calculated using the slope of the log-log plot of $R/S$ vs. $n$ over the entire range of $n$ and are summarized in table \ref{tab1}. 

As shown in the right bottom panels from figure \ref{fig1}, a spectrum of $H$ exponent can be found using a Kernel density estimation. In certain cases, the width of the distribution of $H$ is slightly narrow with a $\sigma^{H}_{Kernel}<0.15$ and, therefore, the data sample can be considered a fractal. However, it is important to emphasize that a further detailed investigation should be needed in this context. For some time series, for instance, Peru's one, the broadness of $H$ can indicate a multifractal signature (see the latter column in table \ref{tab1}). On the other hand, as shown in the left bottom panels, the fluctuations around the straight line are negligible within 3$\sigma$ and, therefore, they can be understood as an observational effect.

As seen in Table \ref{tab1}, all values of these exponents reflect the presence of a persistence of subduction zone statistics, i.e., the values of $H$ are always greater than 0.5, mostly fluctuating around 0.65. In this sense, the values of $H>0.5$ might show that earthquakes in the Ring of Fire are not a purely Gaussian process, on the contrary, there exists a long-term memory associated to the fluctuation dynamics \cite{li}.

Firstly, we verified that only the $b_{s}$-index provides a reasonable adjustment with $H$. This relationship can be observed in Fig. \ref{fig2}, when we split up the data points in two regimes with different slopes. We fit the following law 
\begin{equation}
\label{eq10}
H=\frac{A}{b_{s}}+C,
\end{equation}
over these regimes and found the values of slopes as $A_{\circ}=0.57\pm 0.12$ and $A_{\bullet}=1.11\pm 0.12$, and intercepts $C_{\circ}=0.17\pm 0.11$ and $C_{\bullet}=-0.45\pm 0.11$. We calculate this anti-correlation by using the Spearman ($r_{S}$) and Pearson ($r_{P}$) correlation coefficients and find that $r^{\circ}_{S}=-0.99$, $r^{\circ}_{P}=-0.96$, $r^{\bullet}_{S}=-0.74$ and $r^{\bullet}_{P}=-0.97$ \cite{press}. We observed that high decline law contains the subduction zones with stronger earthquake magnitudes (number 1), while the low decline one is predominantly formed by weaker magnitudes (number 4). In particular, the values of $H$ can be associated to the mechanism which controls the abundance of magnitude and, therefore, the level of activity of earthquakes.

\subsection{Is it possible an a priori determination of the index $H$ from $b_{S}$?}

Inspired by works of Borland \cite{borland}, Sarlis \textit{et al}. \cite{sarlis} and Scherrer \textit{et al}. \cite{scherrer}, and the empirical relation found in previous Section, we investigate a possible theoretical correlation between the indexes $H$ and $b_{s}$ based on the entropic index $q$. 

Borland \cite{borland} introduced a very interesting relation between the $q$-index and the exponent $H$ by relation
\begin{equation}
\label{eq6}
H=\frac{1}{3-q}.
\end{equation}
The author has mentioned that the above relationship is obtained by Langevin equation as a function of the entropic index $q$. This relationship is only valid for $-\infty<q<2$ because of the range of the $H$-index is defined between zero and unity.

The equations (\ref{eq5}) and (\ref{eq6}) suggest that a relation between $H$ and $b_{s}$ can be achieved. After a little of algebra, we found the following expression
\begin{equation}
\label{eq7}
H=0.5+\frac{0.5}{1+b_{s}},
\end{equation}
indicating an anti-correlation between the indexes.

As we can observe in Figure \ref{fig2}, the above equation is not in agreement with the found values of the parameters $A$ and $C$ extracted from empirical relationship (\ref{eq9}). Instead, equation (\ref{eq7}) can be used to estimate the lower and upper values of $H$, but not the their specific ones. According to this equation, $H$ is limited to interval 0.5 and 1.0, considering that $b_{s}=0$ if $q=2$ and $b_{s}\rightarrow \infty$ if $q=1$. Thus, $0.5<H<1.0$ means $1<q<2$, as observed by the values of $q$ shown in table 4 from Scherrer \textit{et al}. \cite{scherrer}.

\section{Final remarks}
We used the well-known Hurst analysis to investigate the behavior a dozen of magnitude series along the Ring of Fire. No doubt, our main result was to estimate the values of the Hurst exponent and to compare with the Gutenberg-Richter indexes. 

From that analysis, we found that there exists a relationship between the Hurst exponent $H$ and the modified Gutenberg-Richter index $b_{s}$ as illustrated by Fig. \ref{fig2}. 

These results revealed that the present sample is consistent with a nonequilibrium state, strongly suggesting that long-term persistence measured by the $H$-index was found among the random variables involved in the physical process that controls seismic activity. The existing of an empirical correlation between the Hurst exponent and the $b_{s}$-index makes it possible to point out toward a promising scenario about the forecasting models of earthquakes due to the high memory which these systems present. Finally, the long-term memory related to the fractal structure of earthquakes will open new ways of analysis in this particular area of research. In addition, we can conclude that dynamics associated with fragment-asperity interactions can be emphasized as a self-affine fractal phenomenon.

Research activities of the Observational Astrophysics and Astrostatistics Board of the Federal University of Cear\'a are supported by CNPq agency. DBdeF also acknowledges financial support by the Brazilian agency CNPq-PQ2 (grant No. 306007/2015-0). RS was also supported by the Brazilian agency CNPq/303613/2015-7.

\end{document}